\documentclass[useAMS,usenatbib]{mn2e}

\usepackage{amssymb}
\usepackage{graphicx}
\usepackage{mathptmx}
\usepackage{amsmath}
\usepackage{color}

\title[Mass-Size Thresholds for Massive Star Formation]{Volume Density Thresholds for Overall Star Formation imply Mass-Size Thresholds for Massive Star Formation}

\author[G. Parmentier et al.]
{Genevi\`eve Parmentier$^{1,2}$\thanks{E-mail: gparm@mpifr-bonn.mpg.de}, 
Jens Kauffmann$^{3}$, 
Thushara Pillai$^{4}$	
and Karl M. Menten$^{1}$   \\
$^{1}$ Max-Planck-Institut f\"ur Radioastronomie, Auf dem H\"ugel 69, D-53121 Bonn, Germany \\
$^{2}$ Argelander-Institut f\"ur Astronomie, Bonn Universit\"at, Auf dem H\"ugel 71, D-53121 Bonn, Germany \\
$^{3}$ NPP Fellow, Jet Propulsion Laboratory, California Institute of Technology, 4800 Oak Grove Drive, Pasadena, CA 91109, USA \\
$^{4}$ California Institute of Technology, MC 249-17, 1200 East California Boulevard, Pasadena, CA 91125, USA}

\begin{document}

\date{Accepted 2011 May 17.  Received 2011 May 13; in original form 2011 February 16
}
\pagerange{\pageref{firstpage}--\pageref{lastpage}} \pubyear{2011}

\maketitle

\label{firstpage}

\begin{abstract}

We aim at understanding the {\it massive} star formation (MSF) limit $m(r) = 870 M_{\odot} (r/pc)^{1.33}$ in the mass-size space of molecular structures  recently proposed by \citet{kau10c}.  As a first step, we build on the property that power-law density profiles for molecular clumps combined with a volume density threshold for {\it overall} star formation naturally leads to mass-radius relations for molecular clumps containing given masses of star-forming gas.  Specifically, we show that the mass $m_{clump}$ and radius $r_{clump}$ of molecular clumps whose density profile slope is $p$ and which contain a mass $m_{th}$ of gas denser than a density threshold $\rho_{th}$ obeys: $m_{clump}=m_{th}^{p/3} \left( \frac{4 \pi \rho_{th} }{3-p} \right)^{(3-p)/3} r_{clump}^{3-p}$.   
In a second step, we use the relation between the mass of embedded-clusters and the mass of their most-massive star to estimate the mimimum mass of star-forming gas needed to form a $10\,M_{\odot}$ star.  Assuming a star formation efficiency of $SFE \simeq 0.30$, this gives $m_{th,crit} \simeq 150 M_{\odot}$.  
In a third step, we demonstrate that, for sensible choices of the clump density index ($p \simeq 1.7$) and of the cluster formation density threshold ($n_{th} \simeq 10^4\,cm^{-3}$), the line of constant $m_{th,crit} \simeq 150 M_{\odot}$ in the mass-radius space of molecular structures equates with the MSF limit for spatial scales larger than 0.3\,pc.  Hence, the observationally inferred MSF limit of Kauffmann \& Pillai is consistent with a threshold in star-forming gas mass beyond which the star-forming gas reservoir is large enough to allow the formation of massive stars.  For radii smaller than 0.3\,pc, the MSF limit is shown to be  consistent with the formation of a $10\,M_{\odot}$ star ($m_{th,crit} \simeq 30 M_{\odot}$ with $SFE \simeq 0.3$) out of its individual pre-stellar core of density threshold $n_{th} \simeq 10^5\,cm^{-3}$.  The inferred density thresholds for the formation of star clusters and individual stars within star clusters match those previously suggested in the literature.    

\end{abstract}


\section{Introduction}
\label{sec:intro}

Massive stars play a crucial role in the evolution of galaxies.  Their ionising radiations and stellar winds carve bubbles into their surrounding interstellar medium (ISM), thereby triggering the formation of new stellar generations \citep{dal09, deh10}, while their explosions as supernovae disperse into the ISM their nucleosynthesis products \citep{heg03}.  Massive forming stars also provide us with the only signatures of star formation in other galaxies.  They are thus important tracers of galaxy star formation rates \citep{cal08}.  The local environment of our Galaxy is an obvious first place where we can secure a firm grasp on their formation conditions, that is, where do they form or not, and why?  On top of that, the mass of the most-massive star in star clusters and the star cluster mass (formally, the cluster stellar mass when it still resides in the molecular clump out of which it forms) are observed to be related to each other \citep{wei06}.  The very nature of that relation remains strongly debated, however.  Is it a {\it physical} one, i.e. is there a cluster-mass-dependent limit upon the maximum stellar mass in a cluster \citep{wei06}?  Or, is it a {\it statistical} relation, i.e. does it merely represent the random-sampling-driven average relationship between the masses of clusters and their most common maximum stellar masses \citep{park07}?  In spite of this uncertainty, there is a wide consensus that the vast majority of massive stars form in star clusters \citep{ll03,park07,wei06}.  The  relation between the embedded-cluster mass, $m_{ecl}$, and the mass of cluster's most-massive star, $m_{*, max}$, demonstrates that the mass of molecular gas forming a star cluster influences the mass of this cluster 's most-massive star since:
\begin{equation}
m_{*, max}=f(m_{ecl})=f(SFE \times m_{CFRg})\,,
\label{eq:m*max}
\end{equation} 
where $m_{CFRg}$ is the gas mass of the cluster-forming region (CFRg) at the onset of star formation and SFE its star formation efficiency.  
Insights in what determines the mass $m_{CFRg}$ of star-forming gas inside individual molecular clumps hold therefore the potential of shedding light on the mass of their respective most-massive star and, eventually, on how massive stars form.  This will also help to model better the impact of stellar feedback upon molecular clumps and molecular clump luminosities, since both are sensitively stellar-mass-dependent.    

\citet{kau10b} recently performed a significant quantitative step towards a better understanding of massive star formation conditions.  Applying the 'dendrogram-technique' developed by \citet{ros08} to column density observations of several solar neighbourhood molecular clouds, they measure the effective radius\footnote{The effective radius is here defined as the radius of the disc whose surface area is identical to that of the contour}, $r$, of many column density contours around column density peaks, as well as the projected mass, $m(r)$, contained within each contour.  In that manner, they obtain a sequence of mass-size measurements over a comprehensive range of spatial scales for each molecular cloud analysed \citep[see fig.~1 in][for an illustration of the method]{kau10a}.  Their combined analysis of the Taurus, Perseus, Ophiuchus and Pipe Nebula molecular clouds -- which all fail at forming massive stars --, and of molecular clumps selected for signposts of massive star formation activity led \citet{kau10c} to conclude that the mass-radius space of molecular structures is characterized by a threshold 
\begin{equation}
m(r) = 870 M_{\odot} (r/pc)^{1.33}
\label{eq:kauflim}  
\end{equation}
for massive star formation (MSF).  That is, the mass-size sequence of molecular clouds devoid of MSF lies {\it below} the aforementioned limiting law \citep[fig.~2a in ][]{kau10b}, while MSF molecular clumps are located {\it above} \citep[fig.~2b in][]{kau10c}.  In other words, MSF demands a mass of molecular gas enclosed within any given projected radius higher than what Eq.~\ref{eq:kauflim} predicts.  In what follows, we refer to Eq.~\ref{eq:kauflim} as the MSF limit.  Note that following the terminology of \citet{kau10c}, a MSF region  comprises HII regions, which have been characterized in literature as regions forming high mass stars, i.e. stars with masses $>8\,M_{\odot}$.

The aim of this contribution is to demonstrate that this limiting law for {\it massive} star formation in the mass-radius space can be explained in terms of a volume density threshold for {\it overall} star formation.  The body of observational evidence in favour of a volume (or number) density threshold for star formation has steadily been growing over the past few years.  On the observational side, there is a tight association between the star formation rate and the mass of high density molecular gas (i.e. hydrogen molecule number densities $n_{\rm H2} \simeq 10^{4-5}\,cm^{-3}$), in both Galactic and extragalactic environments \citep{gao04,wu05, lad10}.  On the theoretical side, \citet{elm07} notes that number densities $n_{\rm H2} \gtrsim 10^5\,cm^{-3}$ enhance microscopic effects able to accelerate star formation significantly (e.g. magnetic diffusion in the molecular gas).  The existence of a number density threshold, $n_{th}$, for star formation also implies that the mean number density of cluster-forming regions is a few times $n_{th}$, the exact factor depending on the clump density profile.  This is in line with the conclusion reached by  \citet{par11a} that constant volume density cluster-forming regions are needed to preserve the shape of the young cluster mass function through the first 50-100\,Myr of cluster evolution.  This effect results from the mass-independent tidal field impact experienced by star clusters dynamically responding to the expulsion of their residual star-forming gas when the mean volume density of their gaseous precursors is constant.  That the shape of the young cluster mass function in the {\it present-day} Universe remains time-invariant for $\simeq 100$\,Myr is suggested by many studies \citep[e.g.][]{cha10}.
[For a more comprehensive summary of the arguments above, see also section 2 of \citet{par11b}].  To investigate how the volume density threshold for {\it overall} star formation and the {\it massive} star formation limit observationally inferred by \citet{kau10b} link to each other is therefore a most timely issue.  

The outline of the paper is as follows.  In Section \ref{sec:mth}, we build a model that relates the MSF limit to the minimum mass of dense star-forming gas needed to form a massive star.  Our conclusions are presented in Section \ref{sec:conclu}.

\section{Massive star formation in dense star-forming gas}
\label{sec:mth}

\subsection{Model}
\label{ssec:mod}
The observed relation between the mass $m_{ecl}$ of an embedded-cluster and the mass $m_{*,max}$ of its most-massive star can be approximated by e.g. the semi-analytical model of \citet[][solid thick line in their fig.~1]{wei06}.  It shows that the formation of stars with masses $m_*$ higher than $\simeq 8$-$10\,M_{\odot}$ requires clusters with a stellar mass $m_{ecl} \gtrsim 100\,M_{\odot}$.  Given that a typical SFE in cluster-forming regions is $\simeq 30\,\%$ \citep{ll03}, the clusters' initial gas masses must be in excess of $m_{th,crit} \simeq 300\,M_{\odot}$ for them to form massive stars.\footnote{Note that an SFE of 30\,\% is also of the order of the formation efficiency required for a star cluster to survive the expulsion of its residual star-forming gas, albeit largely depleted of its initial stellar content  \citep[see fig.~1 in][]{par07}}  Building on the hypothesis that star formation takes place in gas denser than a given number density threshold $n_{th}$, to understand the origin of the MSF limit therefore equates with understanding how the minimal star-forming gas mass, $m_{th,crit}$, relates to the mass and radius of molecular structures.  That is, a star-forming gas mass $m_{th}$ larger than $m_{th,crit}$ leads to the formation of massive stars ($m_{*,max} \ge 10\,M_{\odot}$), while $m_{th}<m_{th,crit}$ is conducive to the formation of clusters less massive than $m_{ecl} \simeq 100\,M_{\odot}$ hence to the formation of low- and intermediate-mass stars only ($m_{*,max} < 10\,M_{\odot}$).  

For the sake of simplicity, let us consider spherically symmetric molecular clumps.  The assumption of spherical symmetry for molecular structures in general is clearly an oversimplification.  However, in the case under scrutiny here, we are interested in dense molecular clumps hosting forming-star-clusters, rather than  the filamentary giant molecular clouds containing them.  As an example, \citet{bel06} mapped in 1.2-mm continuum a large sample of clumps.  They find the mean and median ratios of the full widths at half maximum of their clumps along two perpendicular axes $x$ and $y$, $FWHM_x/FWHM_y$, to be 1.04 and 0.96, respectively, which justifies the assumption of spherical symmetry. 

As for the density profiles of molecular clumps, power-laws have been put forward by various studies, e.g. \citet{hea93}, \citet{hat00}, \citet{beu02}, \citet{fon02} and \citet{mue02}: 
\begin{equation}
\rho _{clump}(s) = k_\rho \, s^{-p}\,,
\label{eq:rho}
\end{equation}
with $s$ the distance from the clump centre and $k_\rho$ a normalizing factor.
We refer to $p$ as the `density index'.  Studies quoted above find $1.5 \lesssim p \lesssim 2$.  In what follows, we will neglect potentially existing radial variations of $p$ within molecular clumps.    

\citet{par11b} (her eq.~3) shows that the mass of gas denser than a volume density threshold, $\rho _{th}$, contained by a spherical clump of mass $m_{clump}$, radius $r_{clump}$ and density index $p$ obeys:  
\begin{equation}
\label{eq:mth}
m_{th} = \left( \frac{3-p}{4 \pi \rho _{th}} \right)^{(3-p)/p} m_{clump}^{3/p} \, r_{clump}^{-3(3-p)/p}\,.
\end{equation}
The mass of gas relevant to star formation is $m_{th}$, rather than the overall clump mass $m_{clump}$.  In what follows, we refer $m_{th}$ as the `mass of star-forming gas' (or $m_{CFRg}=m_{th}$ in Eq.~\ref{eq:m*max}).  Conceptually, molecular clumps in the present paper correspond to picking up a single point along a sequence of mass-radius measurements in e.g. top panel of fig.~2 in \citet{kau10b} (see also our Fig.~\ref{fig:ap}).  

Any convincing model accounting for the MSF limit in the mass-radius plane must address two distinct issues: the slope of the limiting law and its vertical location (equivalently, its intercept).  This issue is easily dealt with using Eq.~\ref{eq:mth}.  

\begin{figure}
\includegraphics[width=\linewidth]{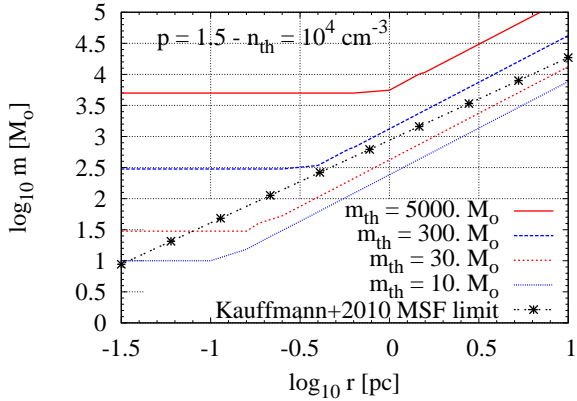}
\caption{Comparison between the observed MSF limit and a grid of iso-$m_{th}$ lines, where $m_{th}$ is the mass of star-forming gas (i.e. $n_{\rm H2} \ge n_{th}$) contained by a clump of mass $r_{clump}$ and $m_{clump}$.  Note that these iso-$m_{th}$ lines are of the same nature as those shown in the top panel of fig.~3 in \citet{par11b}  \label{fig:gridmth} }
\end{figure}

Figure \ref{fig:gridmth} shows the \citet{kau10b} MSF limit (dotted black line with asterisks) along with  lines of constant $m_{th}$, for a density index $p=1.5$ and a density threshold for star formation $n_{th}=10^4cm^{-3}$ \citep[which equates to $\rho_{th} \simeq 700M_{\odot}.pc^{-3}$; ][]{lad10}.  It immediately appears that -- for this sensible choice of parameters -- the observed MSF limit agrees nicely with the line $m_{th} \simeq 300\,M_{\odot}$, at least when $r_{clump} \ge 0.3$\,pc.  $m_{th} \simeq 300\,M_{\odot}$ is precisely the lower bound of the regime for massive star formation.  At this stage, we can already conclude that the hypothesis of a number density threshold for overall star formation and the existence of an MSF limit in the mass-radius space are clearly linked.  Below the MSF limit, star-forming gas masses are lower than $300\,M_{\odot}$, which results in clusters less massive than $\simeq 100\,M_{\odot}$, thus, stars less massive than $\simeq 10\,M_{\odot}$. 

The `break-point' in each iso-$m_{th}$ line (e.g. at $r_{clump} = 1\,pc$ for $m_{th}=5000\,M_{\odot}$) stems from clumps located to the left of the `break-point' having a density at their outer limit at least equal to the assumed threshold $n_{th}$ (i.e. $\rho(r_{clump}) \ge \rho_{th}$).  As a result, $m_{clump}=m_{th}$ for those clumps.  To the right of the break-points, cluster-forming regions (i.e. where $n_{\rm H2} \ge n_{th}$) represent a fraction only of their host-clump volume \citep[see fig.~2 in ][]{par11b}.    

Equation \ref{eq:mth} explains straightforwardly why the slope of an iso-$m_{th}$ line agrees well with that of the MSF limit.  A constant $m_{th}$ leads to $m_{clump} \propto r_{clump}^{(3-p)}$.  For density indices $1.5 \le p \le 2$, the slope $3-p$ of an iso-$m_{th}$ line therefore ranges from $1.5$ to $1$, which brackets nicely the slope of 1.33 estimated by \citet{kau10b} for the MSF limit (see Eq.~\ref{eq:kauflim}).

\begin{figure}
\includegraphics[width=\linewidth]{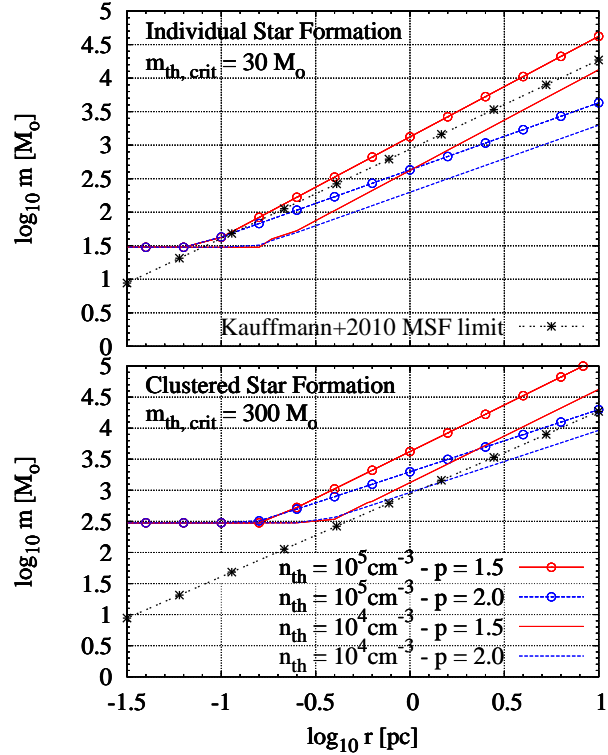}
\caption{How the location of iso-$m_{th}$ lines in the radius-mass space responds to model input parameter variations.  Top and bottom panels consider two distinct masses of star-forming gas, either $m_{th}=30M_{\odot}$ or $m_{th}=300M_{\odot}$.  Assuming SFE$\simeq 0.3$, the first case corresponds to the formation of a $10M_{\odot}$ star out of its individual pres-stellar core, while the second case is relevant to the formation of a $10M_{\odot}$ star within a cluster assuming the relation between the maximum stellar mass and the embedded-cluster mass of \citet{wei06}.  Values for $p$ and $n_{th}$ are quoted in the key. \label{fig:disc}}
\end{figure}

The exact location of the $m_{th,crit}$ line -- our model-proxy to the observed MSF limit -- in the radius-mass space depends on our choice of model parameters, density index $p$ and number density threshold $n_{th}$.  This issue is investigated in Fig.~\ref{fig:disc} with $p=1.5$ or $p=2$, and $n_{th}=10^4cm^{-3}$ or $n_{th}=10^5cm^{-3}$.  
For the sake of completeness, we envisage both clustered star formation (bottom panel), as detailed above, and individual star formation (top panel).  As for the latter, we also assume $SFE \simeq 0.3$ based on the comparison between the core\footnote{We adhere to the following nomenclature: the word `core' refers to the gaseous precursor of an individual star or of a small group of stars, while the term `clump' is designated for regions hosting cluster formation.} mass function and the stellar IMF performed by \citet{alv07} for the Pipe dark cloud.  As a result, the formation of an individual massive star requires a mass of star-forming gas larger than $m_{th, crit} = 30M_{\odot}$.         

The formation of a $10\,M_{\odot}$ star via a clustered mode of star formation requires a larger amount of star-forming gas -- hence a more massive clump -- compared to the formation of an individual star.  As a result, the iso-$m_{th}$ lines move upwards in the bottom panel with respect to the top one.  A higher density threshold for star formation implies that the build-up of a given mass $m_{th}$ of star-forming gas requires a larger amount of molecular gas (since a lower mass fraction of the clump  takes part in the star formation process).  As a result, the $n_{th}=10^5\,cm^{-3}$ lines are located above  their $n_{th}=10^4\,cm^{-3}$ counterparts.  Similarly, a shallower clump density profile is conducive to a smaller fraction of star-forming gas, and $p=1.5$ thus leads to higher clump masses than $p=2$.  Figure \ref{fig:disc} demonstrates that all these sensible parameter values can explain the observed MSF limit inferred by \citet{kau10b}, possibly with the exceptions of the ($n_{th}=10^4\,cm^{-3}$, $p=2$) model for the individual star formation, and the ($n_{th}=10^5\,cm^{-3}$, $p=1.5$) model for the clustered mode of star formation.  The difference between those models and the MSF limit amounts to almost a factor of 10, while in all other cases, the difference is never larger than a factor of three.  We come back to this point in Section \ref{ssec:combined}.

\subsection{Uncertainties on inferring the density threshold}
\label{ssec:unc}
\begin{figure}
\includegraphics[width=\linewidth]{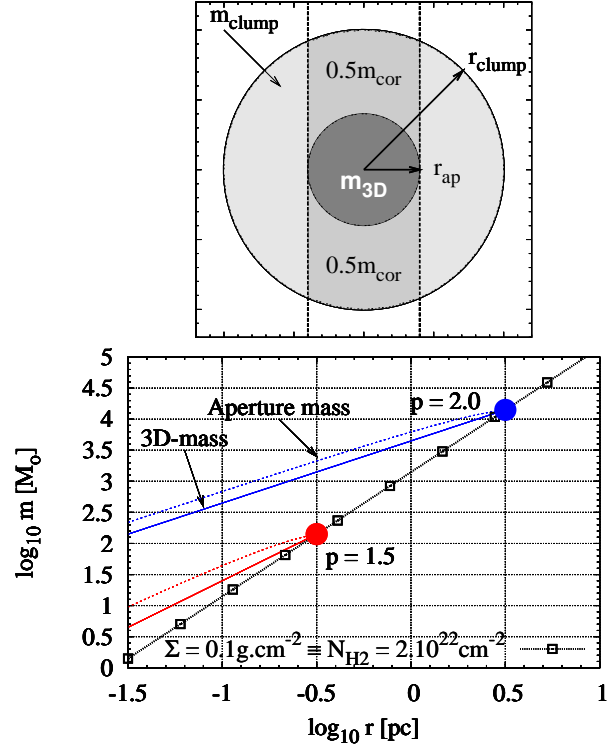}
\caption{Mass-size measurements for two clumps (filled red and blue circles) with distinct density indices ($p=1.5$ or $p=2$).  Solid symbol-free lines depict three-dimensional mass-size measurements, i.e. $m_{3D}(r_{ap})$.  Upper dotted symbol-free lines give the two-dimensional mass-size measurements $m_{2D}(r_{ap})$ by accounting for the clump material located in the foreground and background of the sphere of radius the aperture $r_{ap}$, as illustrated in the top panel where the line of sight is a vertical.  The dotted (black) line with open squares describes another type of clump mass-radius relation, namely, how clump outer radius and total mass are related.  For instance, both clumps depicted here have the same mean column density, $N = 2 \times 10^{22}\,cm^{-2}$.  This type of relation is the main topic of \citet{par11a} and is not to be confused with the radial mass distributions within individual clumps, of relevance for the present contribution     \label{fig:ap} }
\end{figure}

Conversely to what we have done in the previous section, one could infer the density index of molecular clumps and the density threshold for star formation based on the MSF limit.  Since the MSF limit equates with a line of constant $m_{th}$, we can match Eq.~\ref{eq:kauflim} to:
\begin{equation}
m_{clump}=m_{th}^{p/3} \left( \frac{4 \pi \rho_{th} }{3-p} \right)^{(3-p)/3} r_{clump}^{3-p}\,,
\label{eq:mr}
\end{equation}
which is another form of Eq.~\ref{eq:mth}.
This gives $p \simeq 1.7$ and $\rho_{th} \simeq 350\,M_{\odot}.pc^{-3}$, or $n_{th} \simeq 0.5 \times 10^4\,cm^{-3}$.  The inferred density index $p$ is in excellent agreement with measurements of density profile slopes based on e.g. dust continuum emission \citep[e.g.][]{mue02}.  The inferred density threshold $n_{th}$, however, is lower than what is found based on extinction maps and young stellar object census in molecular clouds \citep[$n_{th} \simeq 10^4\,cm^{-3}$, ][]{lad10}, HCN-based studies of molecular gas \citep[$n_{th} \simeq 3 \times 10^4\,cm^{-3}$, ][]{gao04}, and H$^{13}$CO$^+$-mapping of molecular gas \citep[$n_{th} \simeq 10^5\,cm^{-3}$, bottom panel of fig.~3 in][]{par11b}.  The cited observations of the high dipole-moment species, HCN and H$^{13}$CO$^+$, sample regions of high density molecular gas in which the 1-0 transition of these molecules is observed.  Estimates of the threshold density $n_{th}$ for star formation are often chosen to be the critical density of these transitions.\footnote{The critical density of a spectral line transition is defined as the density for which the collisional excitation rate to the transition's upper energy level is equal to its radiative decay rate.} Therefore, the density threshold for star formation inferred from the MSF limit, $n_{th} \simeq 0.5 \times 10^4\,cm^{-3}$, is lower than these line transition densities, as well as lower than the density threshold inferred by \citet{lad10}.  We come back to this point following a discussion of the various  uncertainties affecting the comparison between predicted and observed MSF limits.
 
Molecular clump mapping may underestimate the initial gas mass since a fraction of the gas mass has been 'fed' to recent and ongoing star formation.  An SFE$\simeq 0.3$ implies that observed gas masses $m_{obs}$ should be increased by a factor $\simeq 1.5$ to recover the mass of gas $m_{th}$ at the onset of star formation (since $SFE = m_{ecl}/m_{th} = m_{ecl}/(m_{ecl}+m_{obs})$ and, thus, $m_{th}/m_{obs} = (1-SFE)^{-1}$).  However, this correction factor $\times 1.5$ applies to the limited volume of {\it cluster-forming regions} of mass $m_{th}$.  The global SFE measured over the scale of an entire molecular clump is necessarily lower \citep[local SFE in cluster-forming regions versus global SFE in molecular clumps; see fig.~8 in][]{par11b} and the correction $\times 1.5$ for the gas 'lost' to star-formation becomes an upper limit.  Stellar-feedback-driven clump gas dispersal constitutes another channel through which observed gas masses of molecular clumps may be lower than their pre-star formation contents.  To assess the impact of this effect is well beyond the scope of this paper.  We will assume that, due to star formation and gas dispersal combined, pre-star formation clump masses may be higher than their observed counterparts by at most a factor 2.  In other words, accounting for star formation and gas dispersal implies that a correction factor $\lesssim 2$ should be applied to the intercept of the observed MSF limit prior to comparing it to model outputs.  However, the observed MSF limit is yet affected by another effect -- an overestimating one -- of about the same amplitude, as we now explain.  

Observed masses are two-dimensional masses seen through an aperture corresponding to a contour of constant surface density and, therefore, include material in the foreground and background of the spherical region contained within this aperture.  This effect is depicted in the top panel of Fig.~\ref{fig:ap} with the line-of-sight the vertical.  The projected mass, $m_{2D}(r_{ap})$,  seen through the aperture of radius $r_{ap}$ obeys:
\begin{eqnarray}
m_{2D}(r_{ap}) &=& m_{3D}(r_{ap}) + m_{cor}(r_{ap})             \nonumber \\ 
               &=& 4 \pi k_{\rho} \int_{0}^{r_{ap}} s^{2-p} ds  \nonumber \\ 
&+& 4 \pi k_{\rho} \int_{r_{ap}}^{r_{clump}} \left( 1- \frac{\sqrt{s^2-r_{ap}^2}}{s} \right) s^{2-p} ds\,,
\end{eqnarray}
where the first and second terms on the rhs account for the sphere of radius $r_{ap}$ centered onto the aperture ($m_{3D}(r_{ap})$), and for the correction for the background and foreground material ($m_{cor}(r_{ap})$), respectively.  The bottom panel compares the three-dimensional mass $m_{3D}(r_{ap})=m_{clump}(r_{ap}/r_{clump})^{3-p}$ (solid lines) to its two-dimensional counterpart $m_{2D}(r_{ap})$ (dashed lines) for two clumps of different density indices $p$.  The ratio between both masses depends on the aperture size compared to $r_{clump}$, and on the density index $p$.  It appears that the observed mass overestimates the three-dimensional mass by a factor not exceeding 2.5 (see Kauffmann et al., in prep for a more detailed analysis).  

The uncertainties overestimating (projection effects) and underestimating (star formation and stellar feedback) the MSF limit are therefore of similar amplitude, which limits the uncertainties of a relative comparison between models and observations, as done at the beginning of this section.    

We stress here that the mass-radius relations defined by \citet{kau10a,kau10b} on the one hand, and \citet{par11a} on the other hand have distinct physical significances.  While in \citet{kau10a, kau10b}, the mass-radius relation describes the radial distribution of the mass of molecular gas regions, the mass-radius relation discussed in depth by \citet{par11a} in the framework of the early survival of star clusters in a tidal field refers to the mass-radius relations of {\it populations of clumps}, that is, the relation between the total mass of individual clumps and their radius at their outer edge.  For instance, the two clumps of Fig.~\ref{fig:ap} have identical mean surface densities ($\Sigma \simeq 0.1\,g.cm^{-2}$ or $N_{\rm H2} \simeq 2 \times 10^{22}cm^{-2}$) but different density profiles ($p=1.5$ and $p=2$).  Note that the choice of mass and density index for the two clumps in that figure is arbitrary and entirely chosen for illustrative purposes. 

When working in the framework of the clustered mode of star formation, another uncertainty stems from the maximum stellar mass vs. embedded-cluster mass relation.  Based on the semi-analytical model of \citet{wei06}, we have assumed that $m_{ecl}=100\,M_{\odot}$ leads to $m_{*,max}=10\,M_{\odot}$.  Yet, inspection of the data points most  recently collected by \citet[][right panel of their fig.~3]{wei10} indicates that, on the average, $10\,M_{\odot}$ stars are hosted by clusters $\simeq 50\,M_{\odot}$ in mass.  This decreases our estimate of $m_{th,crit}$ down to $150\,M_{\odot}$ (again with $SFE \simeq 0.3$ in the cluster-forming region) and increases by a factor 2.4 the inferred density threshold, resulting in $n_{th}\simeq 10^4\,cm^{-3}$, in good agreement with \citet{lad10}.          

Finally, we expect our analysis to break down for large enough spatial scales, that is, on scales comprising several clumps since the assumption of spherical symmetry then ceases being valid.  In the Galaxy, our theory thus may not apply beyond radii of several pc.  Note also that whether the observed MSF limit stands for effective radii larger than 10\,pc, or not, is currently not known.

\subsection{How much star-forming gas? A swift graphic approach}
\label{ssec:graph}

\begin{figure}
\includegraphics[width=\linewidth]{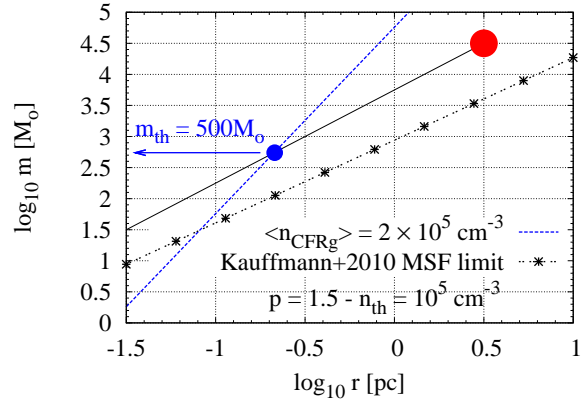}
\caption{How to estimate graphically the star-forming gas mass for a clump with given mass and radius (large filled (red) circle).  $m_{th}$ follows from the intersection between the clump mass radial distribution and  the mean number density $<n_{CFRg}>$ matching given number density threshold $n_{th}$ and density index $p$ (see Eq.~\ref{eq:av_rhoth}).    \label{fig:concept} }
\end{figure}

To estimate the mass $m_{th}$ of star-forming gas of a molecular clump -- hence its ability to form massive stars for a given SFE -- can even be done graphically, without resorting to Eq.~\ref{eq:mth}.  Let us consider a clump of mass $m_{clump} \simeq 3 \times 10^4\,M_{\odot}$, radius $r_{clump} \simeq 3\,pc$ and density index $p=1.5$ (large filled (red) circle in Fig.~\ref{fig:concept}).  The solid (black) line traces the clump three-dimensional mass radial distribution, i.e. $m(s) \propto s^{(3-p)}$.         
The mean density of the star-forming region of mass $m_{th}$ obeys \citep[see eq.~5 in][]{par11b}:
\begin{equation}
<n_{CFRg}> = \frac{3}{3-p} n_{th}\,.
\label{eq:av_rhoth}
\end{equation}
A density index $p=1.5$ is thus conducive to a mean number density $<n_{CFRg}>$ within the star-forming region twice as high as on its outer bound.  This mean density $<n_{CFRg}>= 2 \times 10^5\,cm^{-3}$ is depicted as the (blue) dashed line in Fig.~\ref{fig:concept} (assuming a threshold $n_{th} = 10^5\,cm^{-3}$).  The intersection between the line of constant $<n_{CFRg}>$ and the clump mass radial distribution, indicated by a filled (blue) circle, indicates therefore the radius and the mass corresponding to a mean number density $<n_{CFRg}> = 2 \times 10^5\,cm^{-3}$ hence a limiting density $n_{th} = 10^5\,cm^{-3}$.  In other words, the mass indicated by the small (blue) filled circle defines the mass $m_{th}$ of star-forming gas, that is, $500\,M_{\odot}$.  A reading of the bottom panel of Fig.~\ref{fig:disc} actually indicates that a clump characterized by those mass, radius, density index and number density threshold has $m_{th} \gtrsim 300\,M_{\odot}$.  It is thus possible to estimate graphically the amount of star-forming gas contained by a molecular clump, by searching for the intersection between the clump mass radial distribution and the mean volume density of the gas denser than the threshold $n_{th}$.

\subsection{Individual- and clustered-MSF: a combined approach}
\label{ssec:combined}

Figure \ref{fig:gridmth} shows that the observed MSF limit and the CFRg model $m_{th} \simeq 300\,M_{\odot}$ agree reasonably well with each other when $r_{clump} \ge 0.3\,$pc.  For smaller clumps, the model corresponds to a line of constant mass (i.e. all the clump gas is denser than the density threshold $n_{th}=10^4\,cm^{-3}$), while the observed limit retains its slope of $1.3$ down to spatial scales of a few $0.01$\,pc \citep[see top and middle panels of fig.~2 in ][]{kau10c}.  Top panel of our Fig.~\ref{fig:disc} suggests that the regime $r_{clump} < 0.3\,pc$ can be accounted for by the formation of stars out of their individual density peaks with $n_{th} \simeq 10^5\,cm^{-3}$.   This `individual-star-formation' picture is actually expected if observations look {\it into} a forming-star-cluster, at the spatial scale of individual pre-stellar cores.  

\begin{figure}[t]
\includegraphics[width=\linewidth]{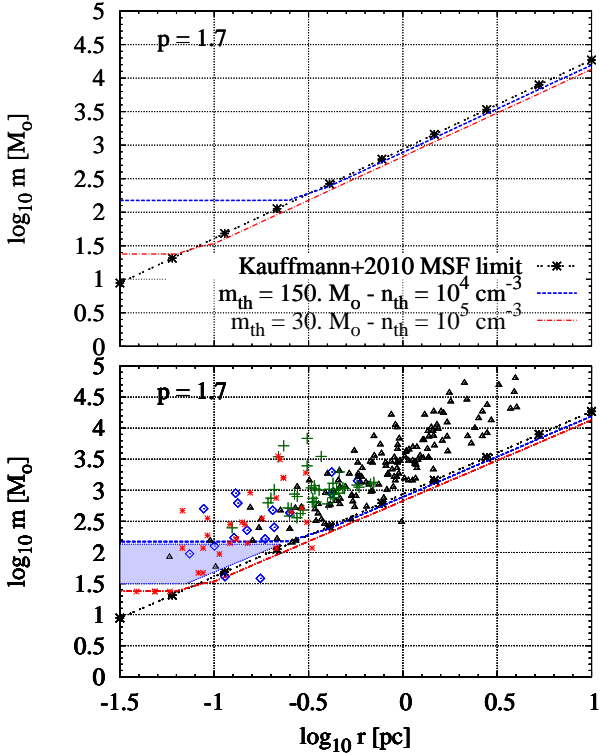}
\caption{
{\it Top panel: } Combination of two MSF models: formation of a 10\,$M_{\odot}$-star out of a density peak with $n_{th}=10^5\,cm^{-3}$ (red dash-dotted line), and formation of a 10\,$M_{\odot}$-star within a CFRg with $n_{th}=10^4\,cm^{-3}$ (blue dashed line).  
The combined model explains the MSF limit down to spatial scales of $\simeq 0.05$\,pc, or clump mass $\simeq 30\,M_{\odot}$, in excellent agreement with the smallest clumps selected for massive star formation activity by \citet{kau10c}.  Note that, for the sake of clarity, the pre-stellar core model (red dash-dotted line) has been shifted by -0.1 in $\log(m)$. 
{\it Bottom panel:} same as top panel, but completed with the data of \citet{kau10c} and the shaded area discussed in the text. \label{fig:comb} }
\end{figure}

Top panel of Fig.~\ref{fig:comb} depicts two models, one for the formation of a star cluster out of a CFRg of mass $m_{th} \simeq 150\,M_{\odot}$ and density threshold $n_{th}=10^4\,cm^{-3}$ (blue dashed line), the other for the formation of a star out of its individual pre-stellar core of mass $m_{th} \simeq 30\,M_{\odot}$ and density threshold $n_{th}=10^5\,cm^{-3}$ (red dash-dotted line).  Assuming $SFE \simeq 0.3$ in both cases \citep{ll03,alv07}, these thresholds in the mass-size space of molecular clumps are consistent with the formation of a 10\,M$_{\odot}$-star on the spatial scale of either a CFRg, or of a pre-stellar core.  Note that since the SFE enters the $m_{clump}$-$r_{clump}$ relation under a power $p/3 \simeq 0.56$ (see Eq.~\ref{eq:mr}), its exact value does not influence the intercept of the models sensitively.  
   
\citet{elm11} confirms that both approaches can be merged into one single coherent picture.  In his model, the average density profile of molecular clumps is described as a decreasing power-law (his equation 5) which, once convolved with a density probability distribution function for supersonically turbulent gas, allows to account for the formation of pre-stellar cores by local turbulence compression all through a clump.  That is, each molecular clump contains many pre-stellar cores \citep[see also][]{mck02}.  Pre-stellar cores correspond to density peaks of at least $n_{th}=10^5\,cm^{-3}$ because stars form fastest where $n_{\rm H2} > 10^5\,cm^{-3}$ \citep[see][his section 3.6]{elm07}.  The mass fraction of clump gas turned into pre-stellar cores by the turbulence is not uniform through a clump, but increases with the clump average density \citep[see fig.~1 in][]{elm11}.  That is, per unit gas mass, pre-stellar cores have a greater probability of forming near the centre for a centrally concentrated clump.  In that respect, our assumption of a number density threshold $n_{th}=10^4\,cm^{-3}$ for clustered star formation constitutes a first-order approximation to \citet{elm11} 's model, i.e. the increased likelihood of forming pre-stellar cores towards molecular clump central regions is replaced by a Heaviside Function $H(n_{\rm H2}-10^4\,cm^{-3})$: $SFE=0$ if $n_{\rm H2} < 10^4\,cm^{-3}$, $SFE \ne 0$ if $n_{\rm H2} \geq 10^4\,cm^{-3}$.      \\   

Whether all massive stars form in star clusters remains disputed \citep[see e.g.][for contrasting points of view]{wit05, park07, gva08, sch08, wei10}.  We note that Fig.~\ref{fig:comb} alone cannot disentangle between the non-clustered and clustered modes of massive star formation.  To illustrate this, let us consider a clump with $r_{clump}=1$\,pc and $m_{clump}=2000\,M_{\odot}$.  The gas masses with $n_{\rm H2} > 10^4\,cm^{-3}$ and $n_{\rm H2} > 10^5\,cm^{-3}$ it contains are 800\,M$_{\odot}$ and 135\,M$_{\odot}$, respectively (using Eq.~\ref{eq:mth} or the graphic approach of Section \ref{ssec:graph}).  With $SFE \simeq 0.3$, this leads to the formation of either an embedded-cluster of mass $\simeq 240\,M_{\odot}$, or a single star of mass $\simeq 45\,M_{\odot}$.  On the average, the most massive star hosted by a  $240\,M_{\odot}$-cluster has a mass $\simeq 15\,M_{\odot}$.  Therefore, both models differ sensitively by the mass of the most massive star formed within the clump.  Due to the high density of molecular clumps, most of the bolometric luminosity given off by the stars forming within their interiors is reprocessed into the infrared.  Besides, the luminosity of main sequence stars depends strongly on their mass (i.e. $L \propto m^{3.5}$).  For the case depicted above, the luminosity of the single-star model thus outshines that of the cluster model by a factor $\simeq 40$.  Combining our model to measures of molecular clump infrared luminosities appear therefore as a promising tool to test whether some massive stars can form outside a cluster environment.   

Bottom panel of Fig.~\ref{fig:comb} reproduces the data collected by \citet{kau10c} in the middle panel of their fig.~2, with the same colour- and symbol-codings.  A handful of molecular clumps are located in the area bounded by the MSF limit and the $m_{th}=30\,M_{\odot}$ and $m_{th}=150\,M_{\odot}$ models (left shaded area in bottom panel of Fig.~\ref{fig:comb}).  Given the small spatial scale ($<0.1\,pc$) and the high mean density ($n_{\rm H2} > 10^5\,cm^{-3}$) of many of these `clumps', the term `core' is better relevant.    
Although those cores might indicate non-clustered massive star formation, they could also belong to a larger --  cluster-forming -- region with a total mass $\gtrsim 150{\rm -} 300\,M_{\odot}$, hence a region expected to form massive stars.  Inspection of the spatial surroundings of the cores in the shaded area of Fig.~\ref{fig:comb} will allow to settle this point.

\section{Conclusions}
\label{sec:conclu}
Building on the model presented in \citet{par11b}, we have demonstrated that power-law density profiles for molecular clumps combined with a volume density threshold for star formation yields mass-size relations for the molecular clumps containing a given amount of star-forming gas hence forming star clusters of a given mass (Fig.~\ref{fig:gridmth}).  We have used this result to explain the massive star formation limit recently inferred by \citet{kau10c}.  

The mass $m_{clump}$, radius $r_{clump}$ and density profile slope $p$ of molecular clumps which contain a mass $m_{th}$ of gas denser than a given density threshold $\rho_{th}$ obey: $m_{clump}=m_{th}^{p/3} \left( \frac{4 \pi \rho_{th} }{3-p} \right)^{(3-p)/3} r_{clump}^{3-p}$ (our Eq~\ref{eq:mr}).  That is, this relation provides the clump mass-size relation corresponding to any given amount $m_{th}$ of star-forming gas for a star formation process driven by a volume density threshold $\rho_{th}$.  
Note that the assumption of a density threshold for star formation implies that formation of star clusters takes place in a limited region of their host molecular clumps \citep[i.e. the cluster-forming region does not necessarily match the volume of the whole clump, see fig.~2 in][]{par11b}.  Our analysis requires very few further assumptions: {\it (i)} molecular clumps hosting forming clusters are spherically symmetric and {\it (ii)} the SFE in cluster-forming regions is mass-independent \citep[as demanded by the mass-independent infant weight-loss of young star clusters, fig.~1 in][]{par07}.  

In the framework of the clustered mode of star formation, we link the mass of a cluster to the mass of its most-massive star with the semi-analytical model of \citet{wei06}, thereby implying that the formation of a $\simeq 10\,M_{\odot}$ star requires a cluster mass of $\simeq 100\,M_{\odot}$.  As star formation efficiencies in cluster-forming regions are expected to be of order $SFE \simeq 0.3$, the minimum mass $m_{th,crit}$ that a cluster-forming region must have to form a massive star is therefore $\simeq 300\,M_{\odot}$.    

Armed with this estimate of $m_{th,crit}$, we have compared the corresponding model for the mass-size sequence of molecular structures (our Eq.~\ref{eq:mr}) with the observed massive star formation limit $m(r) = 870 M_{\odot} (r/pc)^{1.33}$ of \citet{kau10c}.  We derive an estimate of $p \simeq 1.7$ for molecular clump density indices -- which is in excellent agreement with other estimates of the literature --, and a number density threshold for star formation $n_{th} \simeq 0.5 \times 10^4\,cm^{-3}$.  This is lower than what is inferred by e.g. \citet[][$n_{th} \simeq 10^4\,cm^{-3}$]{lad10} based on extinction maps and young stellar object inventories of molecular clouds.  While the difference may be related to uncertainties inherent to our method, we also speculate that this may indicate that $10\,M_{\odot}$ stars can form out of clusters slightly less massive than $100\,M_{\odot}$, as suggested by the star cluster and stellar masses most recently collected by \citet{wei10}.  

The formation of massive stars as members of a star cluster allows to explain the observed MSF limit down to
a spatial scale of 0.3\,pc.  For smaller radii of molecular structures, we show that the MSF limit is consistent with the formation of a $10\,M_{\odot}$ star out of its individual density peak with $n_{th} \simeq 10^5\,cm^{-3}$.  This  is the density of pre-stellar cores predicted by \citet{elm07}.  The observed MSF limit therefore embodies information about the formation of massive stars as star cluster members, and the formation of massive stars out of their individual pre-stellar cores.  In this framework, the density threshold $n_{th} \simeq 10^4\,cm^{-3}$ for clustered star formation probably represents the mean density above which the formation of local density peaks with $n_{th} \simeq 10^5\,cm^{-3}$ is favoured in supersonically turbulent gas.

\section{Acknowledgments}
GP acknowledges support from the Max-Planck-Institut f\"ur Radioastronomie (Bonn) in the form of a Research Fellowship.  JK thanks Di Li, his host at JPL, for making this research possible. This project was supported by an appointment of JK to the NASA Postdoctoral Program at the Jet Propulsion Laboratory, administered by Oak Ridge Associated Universities through a contract with NASA. JK 's research was executed at the Jet Propulsion Laboratory, California Institute of Technology, under a contract with the National Air and Space Administration.  TP acknowledges support from the Combined Array for Research in Millimeter-wave Astronomy (CARMA), which is supported by the National Science Foundation through grant AST 05-40399.  We thank the referee for the original suggestion to combine the individual- and clustered-star formation models into one additional figure.

\end{document}